\documentclass[pra,amsmath,aps,10pt,letterpaper,tightenlines, twocolumn]{revtex4}

\usepackage{bm}
\usepackage{stmaryrd}
\usepackage{amsmath}
\usepackage{calligra}
\usepackage{mathtools}
\usepackage{amsfonts}
\usepackage{amssymb}
\usepackage{textcomp}
\usepackage{graphicx}
\usepackage{caption}
\usepackage[margin=15mm]{geometry}
\usepackage{yfonts}
\usepackage[breaklinks=true,colorlinks=true,linkcolor=blue,urlcolor=blue,citecolor=blue]{hyperref}
\usepackage[lofdepth,lotdepth]{subfig}

\usepackage{color}

\begin{document}

\title{Mutual information as an order parameter for quantum synchronization}
\author{V. Ameri$^1$}
\author{M. Eghbali-Arani$^1$}
\author{A. Mari$^2$}
\author{A. Farace$^2$}
\author{F. Kheirandish$^1$}
\author{V. Giovannetti$^2$}
\author{R. Fazio$^{2,3}$}

\affiliation{$^1$Department of Physics, University of Isfahan, 81746-73441 Isfahan, Iran. \\
$^2$NEST, Scuola Normale Superiore and Istituto Nanoscienze-CNR, I-56127 Pisa, Italy.\\
$^3$Centre for Quantum Technologies, National University of Singapore, 3 Science Drive 2, 117543 Singapore.}

\begin{abstract}
 Spontaneous synchronization is a fundamental phenomenon, important in many theoretical studies and applications. Recently this effect has been analyzed and observed in a number of physical systems close to the quantum mechanical regime. In this work we propose the mutual information as a useful order parameter which can capture the emergence of synchronization in very different contexts, ranging from semi-classical to intrinsically quantum mechanical systems. 
Specifically we first study the synchronization of two coupled Van der Pol oscillators in both classical and quantum regimes and later we consider the synchronization of two qubits inside two coupled optical cavities. In all these contexts, we find that mutual information can be used as an appropriate figure of merit for determining the synchronization phases, independently of the specific details of the system.
\end{abstract}

\maketitle
\section{Introduction}

In 1665 C.\ Huygens first observed that two pendulum clocks mounted on the same support tend to oscillate in a synchronous way~\cite{huygens5letterbis}. This is a particular instance of a rather widespread phenomenon called spontaneous synchronization \cite{pikovsky,strogatz}. In basic terms, two or more classical systems spontaneously synchronize when the amplitude and/or the phase of their individual phase-space trajectories lock together due to some mutual coupling, in the complete absence of any external reference signal. This behavior has been observed in a large variety of biological, chemical, physical and social contexts~\cite{strogatz} and has become a well-understood feature of non-linear classical systems~\cite{pikovsky}.

Recently there have been considerable efforts to extend the concept of synchronization to quantum systems, where the notion of deterministic trajectories in phase space is no longer meaningful. Most approaches deal with continuous variable systems that can still be conveniently described by quasi-distributions in phase space, retaining some link with the classical theory. 

Synchronization has been characterized by looking at the localization of the Wigner function or by comparing the local frequency spectra of optical \cite{Lee2013b} and mechanical resonators \cite{Heinrich2011a,Ludwig2013a, Mari2013a}. Along the same research line,  extrapolating them from the concepts of complete and phase synchronization of classical models, quantitative measures of synchronization for continuous variable quantum systems have been recently proposed \cite{Mari2013a}, highlighting the fundamental limits imposed by the Heisenberg uncertainty principle. 
A paradigmatic example in this context is given by the Van der Pol (VdP) oscillator, i.e. the simplest model of a non-linear resonator characterized by self-sustained oscillations. Spontaneous synchronization between two quantum VdP oscillators, both coherently~\cite{Lee2013a} and dissipatively~\cite{Lee2014a,Walter2014b}~ coupled, or many VdP oscillators~\cite{Lee2013a,Lee2014a} has been characterized  as well. Phase locking of a single VdP resonator with an external drive was studied in \cite{Walter2014a}.
These systems are very promising from an experimental point of view, because the first observations of classical synchronization of nano- and micromechanical oscillators have been recently reported~\cite{Shim2007,Zhang2012,Bagheri2013a,Matheny2014} and the quantum regime is not far from current technological capabilities. 

Quantum networks of two~\cite{Giorgi2012a} or more~\cite{Manzano2013a} coupled linear (harmonic) oscillators have also been investigated. Here the system usually reaches a stationary static configuration, and synchronization can be observed in the initial transient regime where it can be characterized by the dynamics of local observables. Moreover, in this case the emergence of synchronization has been related to the presence of slow-decaying quantum correlations, in the specific form of quantum mutual information and quantum discord~\cite{Modi2012a}. This approach has also been extended to finite dimensional quantum systems such as two dissipatively coupled spins~\cite{Giorgi2013a}. Finally, very recently, synchronization between coupled quantum many-body systems~\cite{Qiu2014a} has been studied and possible tests with bosonic ultra-cold atomic clouds have been proposed \cite{tieri}.

Several aspects are still not completely explored. How can we universally define and quantify synchronization for finite dimensional systems? Is there a relationship between synchronization and general quantum correlations? With the present work, we attempt to give some answers to these questions.  Specifically we suggest the use of  quantum mutual information as an order parameter for quantitatively  determining the synchronized phase of arbitrary quantum systems. The advantage of this information-based approach is that it applies both to semi-classical continuous variable systems where quantum fluctuations add noise around classical trajectories, but also to deeply quantum systems ({\it e.g.} qubits) where the idea of synchronization cannot even be visualized in terms of a classical analogue. 

The paper is organized as follows: in the first section we review the concept of mutual information between two quantum systems and propose its use as an order parameter for synchronization. We then apply this approach to the prototypical scenario of two coupled quantum Van der Pol oscillators, evolving in the semi-classical and in the quantum regime. We show that the mutual information gives a good characterization of spontaneous synchronization, in qualitative agreement with the phase-space synchronization measure introduced in Ref.\ \cite{Mari2013a}. As a second step, we study the dynamics of an intrinsically quantum system consisting of two optically coupled qubits, where typical features of synchronization (finite threshold, Arnold tongue, {\it etc.}) are characterized in terms of the mutual information. Finally we analyze the interplay between classical and quantum correlations in the emergence of synchronization.

\section{Mutual information and syncrhonization}
Given a quantum state $\rho$ composed of two subsystems with reduced density matrices $\rho_A={\rm Tr}_B (\rho)$ and $\rho_B={\rm Tr}_A (\rho)$ respectively, the quantum mutual information
is defined as 
\begin{eqnarray}\label{mi}
I= S(\rho_A)+S(\rho_B)-S(\rho),
\end{eqnarray}
where $S(\rho)=-{\rm Tr}[ \rho \log(\rho)]$ is the Von Neumann entropy.
In classical information theory,   mutual information is a measure of correlations between two random variables $A$ and $B$. Operationally, it quantifies how much the knowledge of the variable $A$ gives information about the variable $B$. Eq. \eqref{mi} is the direct generalization of this quantity to systems described by quantum states. 

The idea behind this work is that synchronized systems should converge to a steady state having large mutual information. In order to better understand this relationship, it is convenient to first consider a classical example. Imagine to have an ensemble of pendulum clocks which are weakly mechanically coupled in such a way that they spontaneously tend to synchronize. If the clocks are not perfect, after a long time, the information about the position of each clock-hand is completely lost (high local entropy). However, since the clocks are synchronized, if we knew the state of one clock we could completely determine the state of the ensemble (low global entropy). From an information theoretic perspective, this scenario corresponds exactly to a system possessing large mutual information.

The advantage of this information-based approach, is that it can be straightforwardly extended to quantum systems simply by replacing the Shannon entropy with the Von Neumann entropy. Basically one can use the expression \eqref{mi} as an order parameter which could signal the presence of a synchronized phase. 
In this way one can study the synchronization of deeply quantum systems such as qubits, where any semi-classical interpretation of this effect is hardly applicable. For example, while for two mechanical resonators one could try to define quantum synchronization extending the idea of ``two systems converging to equal phase-space trajectories"  to quantum operators (see e.g.~\cite{Mari2013a}), for two qubits this classically motivated approach cannot be used. Nonetheless  mutual information is still a well defined quantity and  this  fact allows us to extend the notion of synchronization to quantum system of arbitrary nature: semi-classical or deeply quantum, continuous variable or discrete variable, {\it etc.}. 

 An enhancement of mutual information during the synchronized dynamics of quantum systems has been already observed in some previous works \cite{Giorgi2012a, Manzano2013a, Giorgi2013a}. The aim of our contribution is to further investigate this link and, more precisely, to underline the universality of mutual information as a proper order parameter for synchronization. The main message that we would like to convey is that the link between synchronization and mutual information is not accidental but, in fact, it can be considered as a kind of ``definition'' of synchronization from an information theory perspective.

In order to move smoothly from the classical towards the quantum regime, here we begin our analysis by first considering two VdP oscillators and, eventually, we will focus on the synchronization of two qubits.

\section{Synchronization of two Van der Pol oscillators}

The VdP oscillator was originally proposed by Balthasar van der Pol in 1920 \cite{van1934nonlinear}. This is basically a model of a harmonic oscillator with additional non-linear terms in the equations of motion and has been successfully used to describe a variety of systems possessing a cyclic behavior. A characteristic feature of the VdP oscillator is the existence of periodic steady state solutions (limit cycles) even when the system is not driven by a time dependent force.  This feature, common in non-linear systems, typically gives rise to synchronization phenomena among different limit cycles associated to  two or more VdP resonators  \cite{strogatz, pikovsky}. For this reason the quantum version of the VdP oscillator represents a perfect candidate for understanding the analogies and the differences between classical and quantum syncrhonizaiton.  Indeed several  theoretical studies have been recently performed~\cite{Lee2013a,Lee2014a,Walter2014b,Walter2014a} based on this approach.

\begin{figure}
\includegraphics[width= \columnwidth]{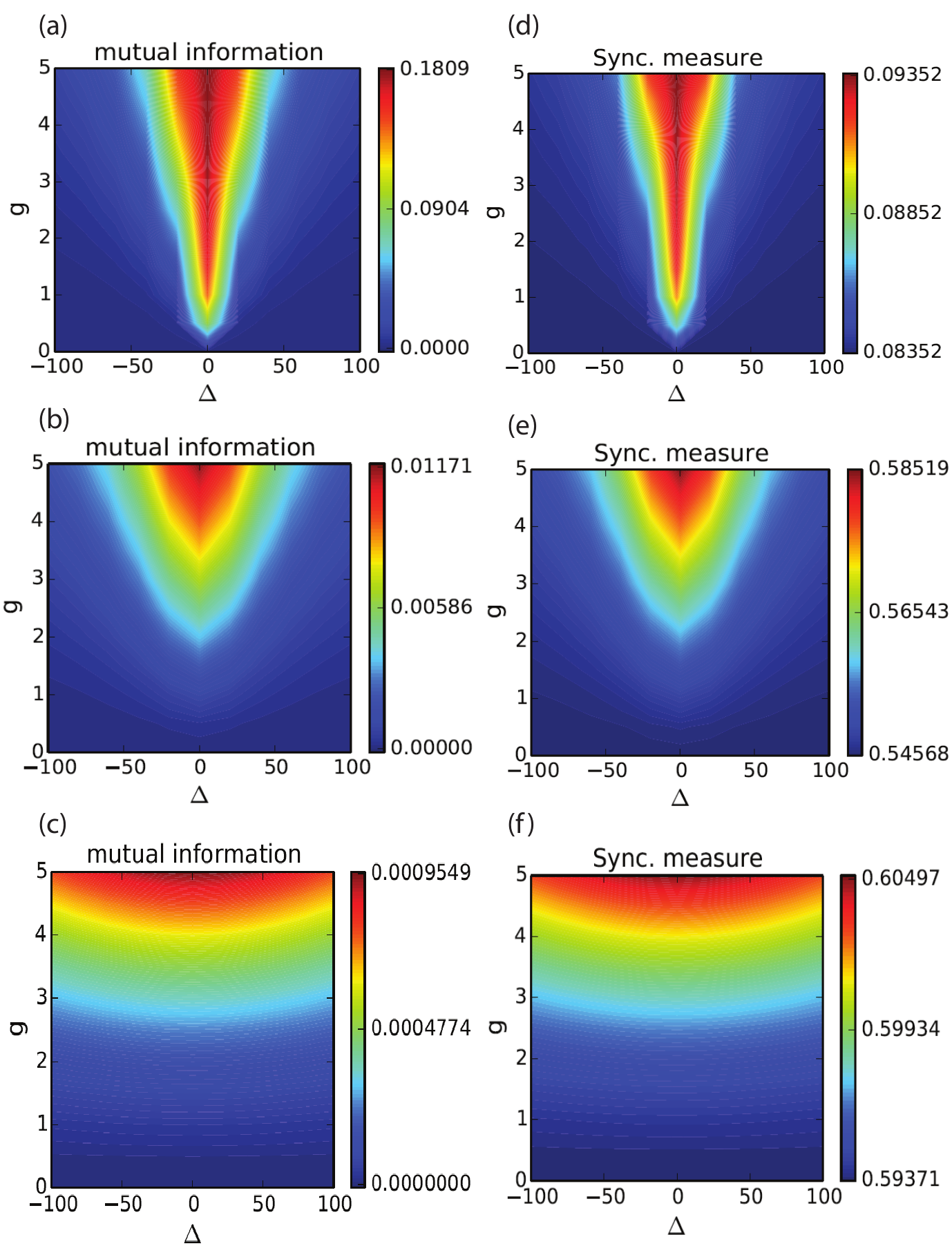}
 \caption{Synchronization analysis for the steady state of the two VdPs model~(\ref{master}): (a,b,c) Mutual information, (d,e,f) Semi-classical synchronization measure defined in Eq.\eqref{measure}. The quantumness parameter:  is for  (a,d)  $\kappa_2/\kappa_1=0.1$, for (b,e)  $\kappa_2/\kappa_1=10$ and for  (c,f) $\kappa_2/\kappa_1=100$.  $\Delta$ is the detuning of VdPs, and $g$ is the coupling strength. All parameters are in units of $\kappa_1$.}\label{vdp}
\end{figure}

The quantum mechanical dynamics of two coupled VdP oscillators can be described by the following master equation \cite{Lee2013a}:

\begin{eqnarray}\label{master}
\frac{d\rho}{dt}=-i\left[ \rho , H \right] +\sum_{i=1}^2\kappa_1 \left( 2a_i^\dagger\rho a_i-a_ia_i^\dagger\rho - \rho  a_ia_i^\dagger\right) \\ \nonumber +\kappa_2\left( 2a_i^2\rho a_i^{\dagger 2}-a_i^{\dagger 2}a_i^2\rho-\rho a_i^{\dagger 2}a_i^2\right), 
\end{eqnarray}
where we set $\hbar=1$, $a_1$ and $a_2$ are the bosonic annihilation operators of each VdP resonator,  $2\kappa_1$ and  $2\kappa_2$ are the rate of gaining one phonon and of losing two phonons respectively. The Hamiltonian is 
\begin{equation}
H=\omega_1{a_1}^\dagger a_1+\omega_2{a_2}^\dagger a_2+g({a_1}^\dagger a_2+{a_2}^\dagger a_1),
\end{equation}
where $\omega_j$ are the natural frequencies of the two oscillators and $g$ is a weak coupling constant which will be responsible for the development of synchronization. 

As shown in~\cite{Lee2013a}, by changing the ratio of the dissipative rates $\kappa_1$ and $\kappa_2$  one can easily interpolate from a classical ($\kappa_2/\kappa_1\ll1$) to a quantum ($\kappa_2/\kappa_1\gg1$) limit.
Indeed for $\kappa_2/\kappa_1\ll1$ the VdP oscillators develop semi-classical limit cycles (large amplitudes and low noise), on the other hand for $\kappa_2/\kappa_1\gg1$ the VdP oscillators have such small amplitudes that the steady state is essentially dominated by quantum fluctuations.
For both regimes we evaluated the mutual information of the steady state with respect to the coupling $g$ and to the detuning between the two VdP oscillators $\Delta=\omega_2-\omega_1$. In Fig.\ref{vdp} one can clearly recognize the typical Arnold's tongue \cite{strogatz, pikovsky} (``V" shape) of the synchronized phase and also the existence of a clear threshold for the parameter $g$, below which synchronization does not happen. Moreover, as already noticed in several works \cite{Lee2013b, Walter2014a}, moving from the classical to the quantum regime the level of synchronization is reduced and the typical Arnold tongue is smoothed due to the presence large quantum fluctuations.  

In order to justify the use of  mutual information as a valid order parameter, we also compare it with a semi-classical measure of complete-synchronization introduced in~\cite{Mari2013a}. This is defined as 
\begin{eqnarray}\label{measure}
S_c=\left\langle  p_-^2+q_-^2 \right\rangle  ^{-1} ,
\end{eqnarray}
where $p_-=(p_2-p_1)/\sqrt{2}$, $q_-=(q_2-q_1)/\sqrt{2}$, and $q_i=(a_i+a_i^\dag)/\sqrt{2}$, $p_i=i(a_i^\dag-a_i)/\sqrt{2}$, are the dimensionless position and momentum operators of the VdP oscillators. 

The interesting result here is that both the synchronization measure \eqref{measure} and the mutual information have the same qualitative behavior, justifying our idea of using mutual information as an order parameter.
Actually, due to the small amount of phonons in the system, the semi-classical quantity \eqref{measure} is significantly non-zero even when the system is not synchronized (this effect is negligible only in the large energy regime). This fact may be considered an unwanted feature for a well behaving order parameter. On the contrary,  mutual information does not suffer from this problem and, even in the deeply quantum regime, the quantity is zero outside  synchronization region. 

As a side-remark we comment that in all the cases considered in Fig. \ref{vdp}, we did not find entanglement even in the presence of synchronization. This fact suggests that there is not a one-to-one correspondence between entanglement and synchronization, even if there are cases in which this relation is present, as recently reported in Ref.~\cite{Lee2014a}.

\section{Synchronization of two qubits}

In the previous section we studied the synchronization of two VdP resonators. Now we consider the synchronization of two qubits which are coupled by optical radiation. As we are going to show, even in this intrinsically quantum case, mutual information can be used as an order parameter for synchronization.

\begin{figure}
\includegraphics[width= \columnwidth]{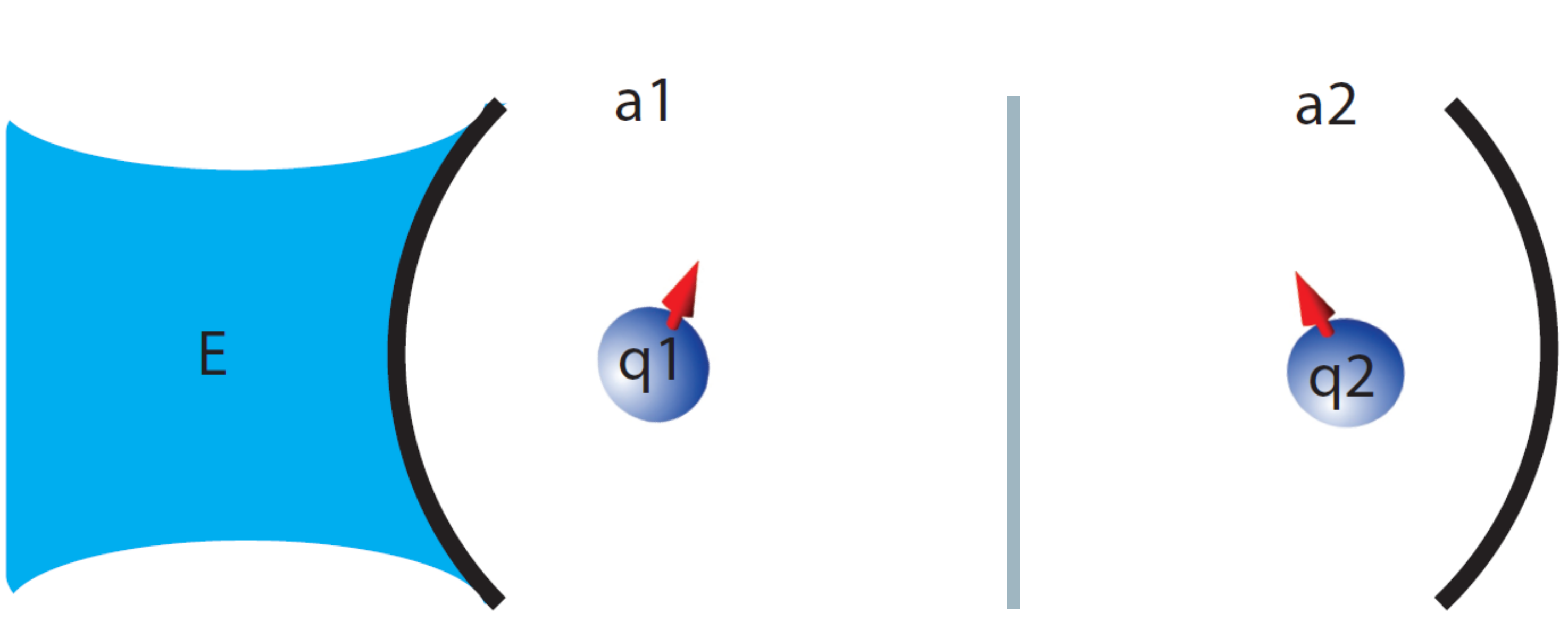}
  \caption{$a_1, a_2$ and $q_1, q_2$ are the optical modes and qubits in the first and second cavities, respectively. $E$ is a driving laser amplitude which is applied to the first cavity. Photons can coherently hop from one cavity to the other with a rate $g$. }\label{qubit}
\end{figure}

We assume that the two qubits are placed in two coupled optical cavities where only the first cavity is driven by a laser, while the second one is populated by the photons leaking from the first cavity. The setup is described in Fig.\ \ref{qubit}  and the corresponding Hamiltonian is the following

\begin{eqnarray}
H &=& \omega_1 a_1^\dag a_1 + \omega_2 a_2^\dag a_2 + \omega_1 \sigma_{z1} + \omega_2 \sigma_{z2} + E(a_1^\dag + a_1) \\ \nonumber &+&g(a_1^\dag a_2 +a_2^\dag a_1) 
+\mu  (a_1+a_1^\dag) \sigma_{x1}  + \mu (a_2+a_2^\dag )\sigma_{x2} ,
\end{eqnarray}
where $ \hbar=1$, $E$ determines the strength of the external driving on the first cavity and  $g$ and $\mu$ are the optical coupling constant and the qubit-field coupling constant respectively. We assume that each cavity is resonant with its own internal qubit, while the detuning  $\Delta=\omega_2-\omega_1$ between the characteristic frequencies of the two qubits can be nonzero.
We also take into account the dissipation of both optical cavities into the environment while, for simplicity, we neglect the direct decoherence of the qubits.
The corresponding master equation is then 
\begin{eqnarray}
\frac{d\rho}{dt}=-i\left[ \rho , H \right] +\sum_{i=1}^2\kappa \left( 2a_i\rho a_i^\dagger-a_i^\dagger a_i\rho - \rho  a_i^\dagger a_i\right).
\end{eqnarray}

In order to study the emergence of synchronization, we compute the mutual information between the two qubits on the steady state of the system as a function
of the detuning $\Delta$ and of the optical coupling constant $g$. 
Similarly to the previous case involving VdP oscillators, the mutual information in non-zero in a parameter region with the characteristic shape of an  Arnold tongue. Moreover also in this case we observe a threshold value of the coupling $g$, below which, mutual information is negligible. These peculiar features justify the interpretation of such correlated phase as a quantum synchronization effect.

\begin{figure}
\includegraphics[width= 0.75  \columnwidth]{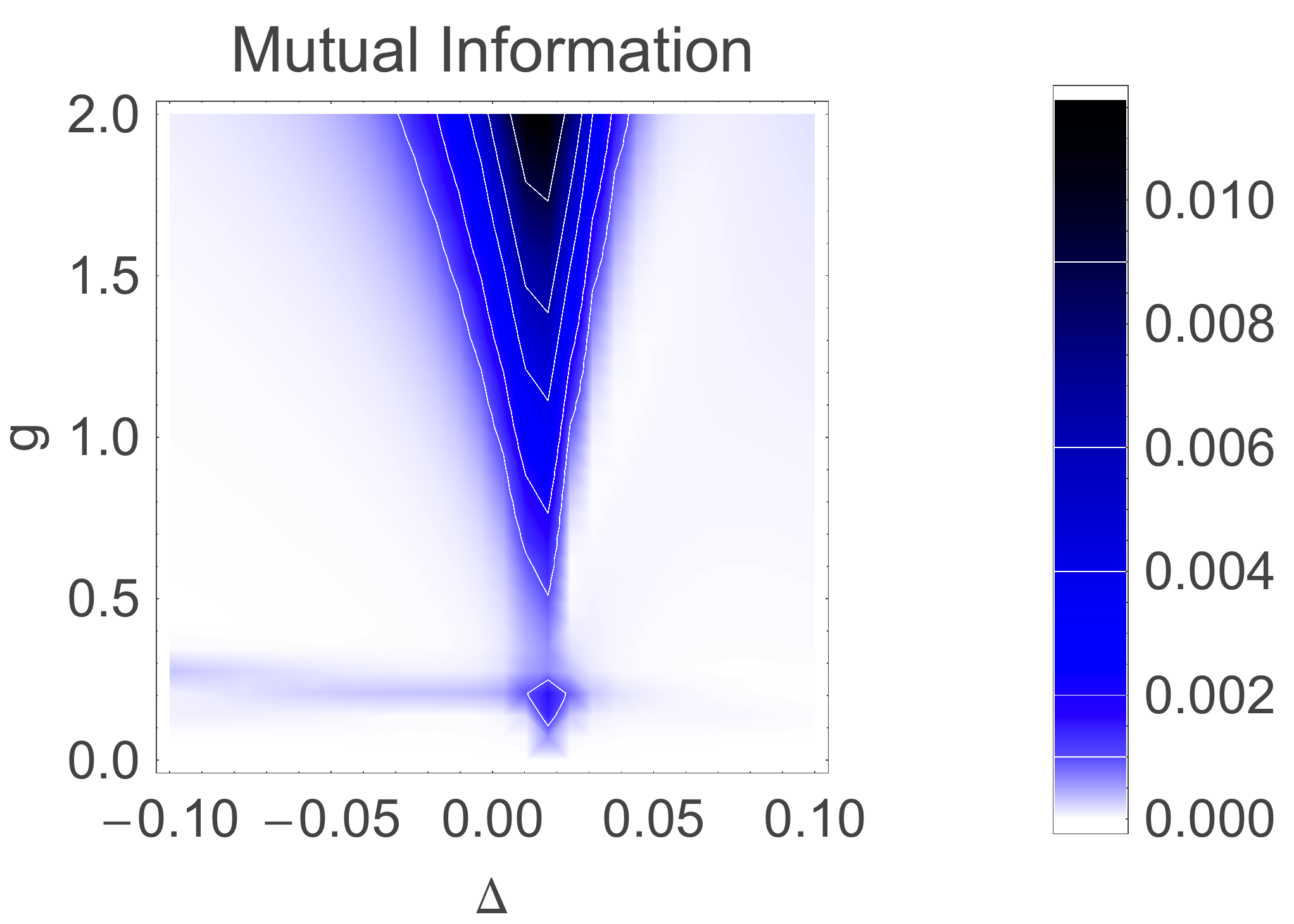}
\caption{Mutual information of the steady state of the two qubits model as a function of the optical coupling constant $g$ and detuning $\Delta$. The other parameters are $\omega_1 =10$, $\kappa=0.05$, $E=3$, and  $\mu=1$.  }\label{mu100}
\end{figure}

Mutual information measures all correlations existing between the qubits and then one could doubt that, despite the strong analogy, the Arnold tongue phase shown in Fig.\ \ref{mu100} could be unrelated to synchronization. For this reason, in order to validate our interpretation, we compare the value of mutual information with a model-specific measure of phase-locking. 
In the previous section about VdP oscillators,  we used the semiclassical measure based on position and momentum quadratures given in Eq.\ \eqref{measure}. In this case instead there is not a classical analogue of the system and, since we deal with two qubits,  Eq.\ \eqref{measure} cannot be applied. 
Then, to visualize the the phase-locking between the qubits, we need to use a different approach (similar to \cite{Giorgi2013a}) which is based on the transient dynamics happening before the system reaches the steady state.  In this initial transient, the expectation values of the operators are time dependent and present Rabi-like oscillations. In our specific model, we found more convenient to focus on the $x$ component of each qubit in the Bloch sphere 
$\langle \sigma_{x,i} \rangle$, $i=1,2$. The simplest model for the oscillations of this quantity is given by
\begin{equation}\label{phosc}
\left\langle \sigma_{x,i} (t) \right\rangle  = \bar{\sigma}_{x,i}(t) \sin[\phi_i(t)]   
\end{equation}
where $\bar{\sigma}_{x,i}(t)$ are slowly varying amplitudes and $\phi_i(t)$ are the oscillation phases ({\it i.e.} we assume $\frac{d}{dt} \bar{\sigma}_{x,i}(t) \ll
\frac{d}{dt} \phi_i(t)$).
Defining the quantity 
\begin{equation}\label{phl}
s_p(t)= \cos(\phi_2(t) - \phi_1(t))
\end{equation}
we get an estimate of the relative phase between the qubits. For example if $s_p\simeq 1$ the qubits are phase-locked, if $s_p\simeq -1$ the
oscillations are anti-phase locked, while if $s_p$ is not stable around a constant value we can conclude that the system is not synchronized.

\begin{figure}
\includegraphics[width= \columnwidth]{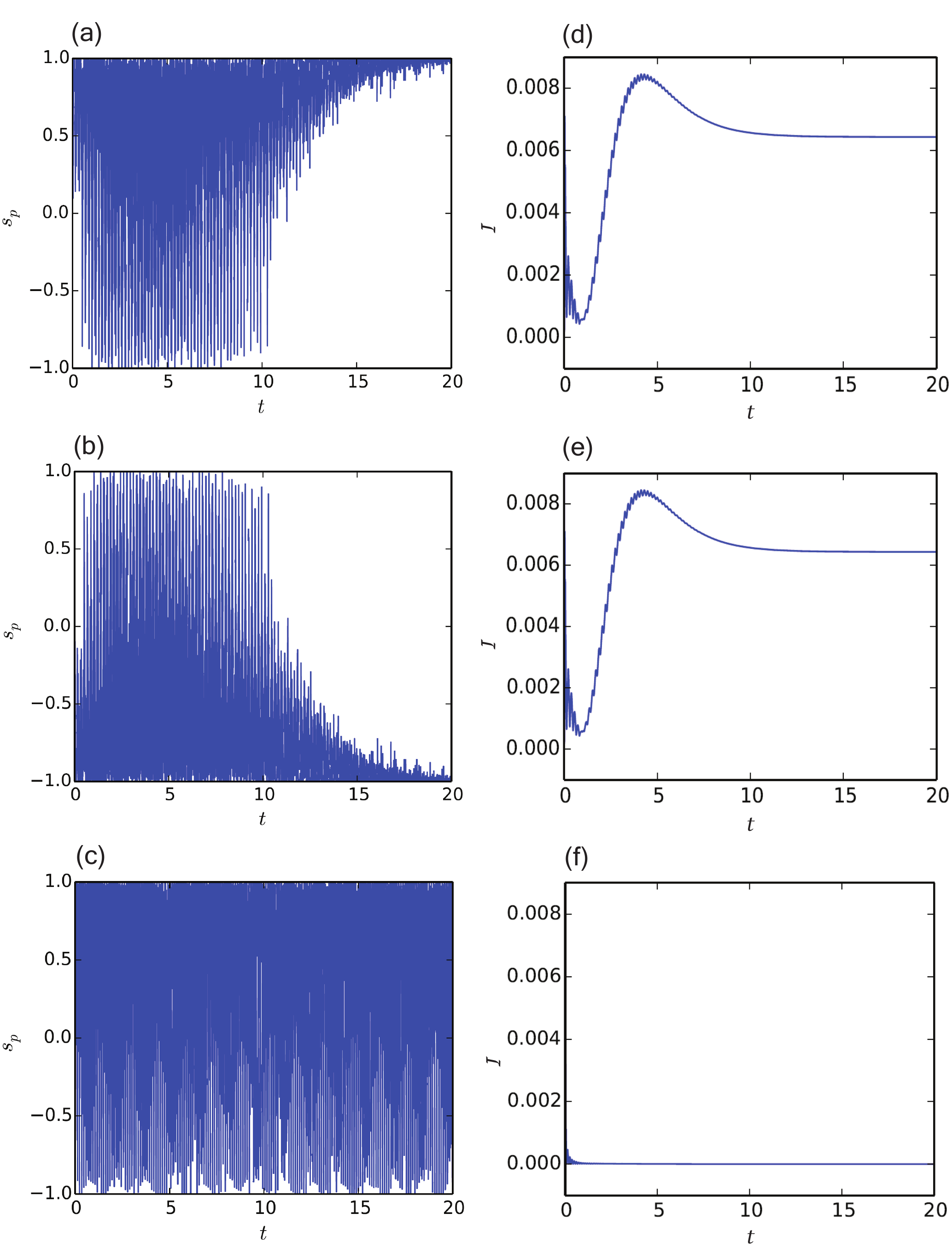}
  \caption{(a,b,c) Simulation of phase locking measure $s_p$ defined in Eq.\eqref{phl} as a function of time (in units of $\tau = 2\pi/\omega_1$ ). (d,e,f) Mutual information of qubits as a function of time (in units of $\tau$). The parameters are (a,d):$ \omega_1=10, \omega_2=10, g=0.5, \mu=1 ,\kappa=0.1$, $E=3$, (b,e):$\omega_1=10, \omega_2=10, g=-0.5, \mu=1 ,\kappa=0.1$ , $E=3$, (c,f):$\omega_1=10, \omega_2=20, g=0.5, \mu=1, \kappa=0.1$, $E=3$.}
   \label{muqubit} 
\end{figure}

In Fig.\ref{muqubit} (a,b) we plot $s_p(t)$ for different values of the parameters and we compare it with the behavior of mutual information. 
Clearly, one can observe that when $\Delta=\omega_2-\omega_1\simeq 0$ the oscillations become phase locked and mutual information assumes nonzero values. On the contrary, when  $\Delta$ is too large, $s_p(t)$ oscillates without any phase-locking effect and correspondingly the mutual information is negligible.  These results justify our initial interpretation of the mutual information phase diagram (Fig.~\ref{mu100}) as a signature of quantum synchronization. 

We remark that the advantage of mutual information with respect to other model-based measures of synchronization lies in its universality and in the possibility of using it as a steady state order parameter (as done in Fig.~\ref{mu100}). Indeed, while a specific measure like $s_p$ loses its meaning when all the initial Rabi oscillations are damped to equilibrium,  mutual information is still able to reveal the presence of synchronization hidden in the cross
correlations between the two subsystems.

\section{Quantum correlations and synchronization}
The use of the mutual information as an order parameter for synchronization allows us to shed more light on the interplay between synchronization and correlations. Quantum mutual information describes the total amount of correlations in a system and this total can be further divided into classical and quantum correlations~\cite{Henderson2001a,Ollivier2001a}. 

Classical correlations can be interpreted as the information gain about one subsystem as a result of a local measurement on the other. In classical information theory this quantity is defined as $I(AB) = H(B) - H(B|A)$, where $H(X)$ is the Shannon entropy of the random variable $X$ and $H(B|A)$ is the conditional entropy of $B$ given $A$. This expression is equivalent to the classical mutual information $I(AB) = H(B) + H(A) - H(AB)$ via Bayes rule. In quantum information theory, there are many different measurements that can be performed on a system and measurements generally disturb the quantum state. Therefore, classical information is defined by taking the maximum over all possible measurements and reads
\begin{equation}\label{class}
I^{C}_{A}(\rho_{{AB}}) = \max_{\{E_{A}\}} \left( S(\rho_{B}) - S(B| \{E_{a}\}) 	\right),
\end{equation}
where $\{ E_{a}\}$ are the elements of a positive operator valued measurement (POVM) on $A$, $S(B| \{E_{a}\}) = \sum p_{a} S\left( \frac{ \rm{Tr}_{A} \left[ E_{a} \; \rho_{AB} \right]}{p_a} \right)$ is the average Von Neumann entropy of the conditional state of $B$ and $p_{a} = \rm{Tr} \left[ E_{a} \; \rho_{AB} \right]$ is the probability of getting outcome $a$.

The difference between the mutual information and the classical information gives the so called quantum-discord~\cite{Henderson2001a,Ollivier2001a}, which measures the genuinely quantum correlations in the system.
\begin{eqnarray}\label{disc}
D_{A}= I - I^C_{A} = S(\rho_A) - S(\rho) + \min_{\{E_{A}\}} S(B| \{E_{a}\}),
\end{eqnarray}

Analogous quantities can be defined if the measurement is performed on subsystem $B$. In general, the classical information and the discord are asymmetric (i.e. $D_{A} \neq D_{B}$), but for the cases examined in this paper the behavior of the two alternatives is both qualitatively and quantitatively similar. 

In the following we compute the classical information and the discord for our two-qubit system. In principle, as prescribed by eqs.~\eqref{class} and~\eqref{disc}, we would have to face the challenging task of optimizing over all POVMs. However, for qubits, several simplifications are at hand. First of all, it was shown in~\cite{Hamieh2004a} that the optimal POVMs must be extremal, i.e. they cannot be written as a convex combination of other POVMs. Moreover, extremal POVMs for qubits can only have 2, 3 or 4 projectors as their elements~\cite{DAriano2005a}. Finally, there are strong numerical evidences that optimization over POVMs with 2 (orthogonal) projectors yields numbers that can be safely considered correct (the error being of the order of $10^{-4}$)~\cite{Galve2011a}. We adopt this approximation to reduce the computational complexity.

\begin{figure}
\includegraphics[width= 0.75 \columnwidth]{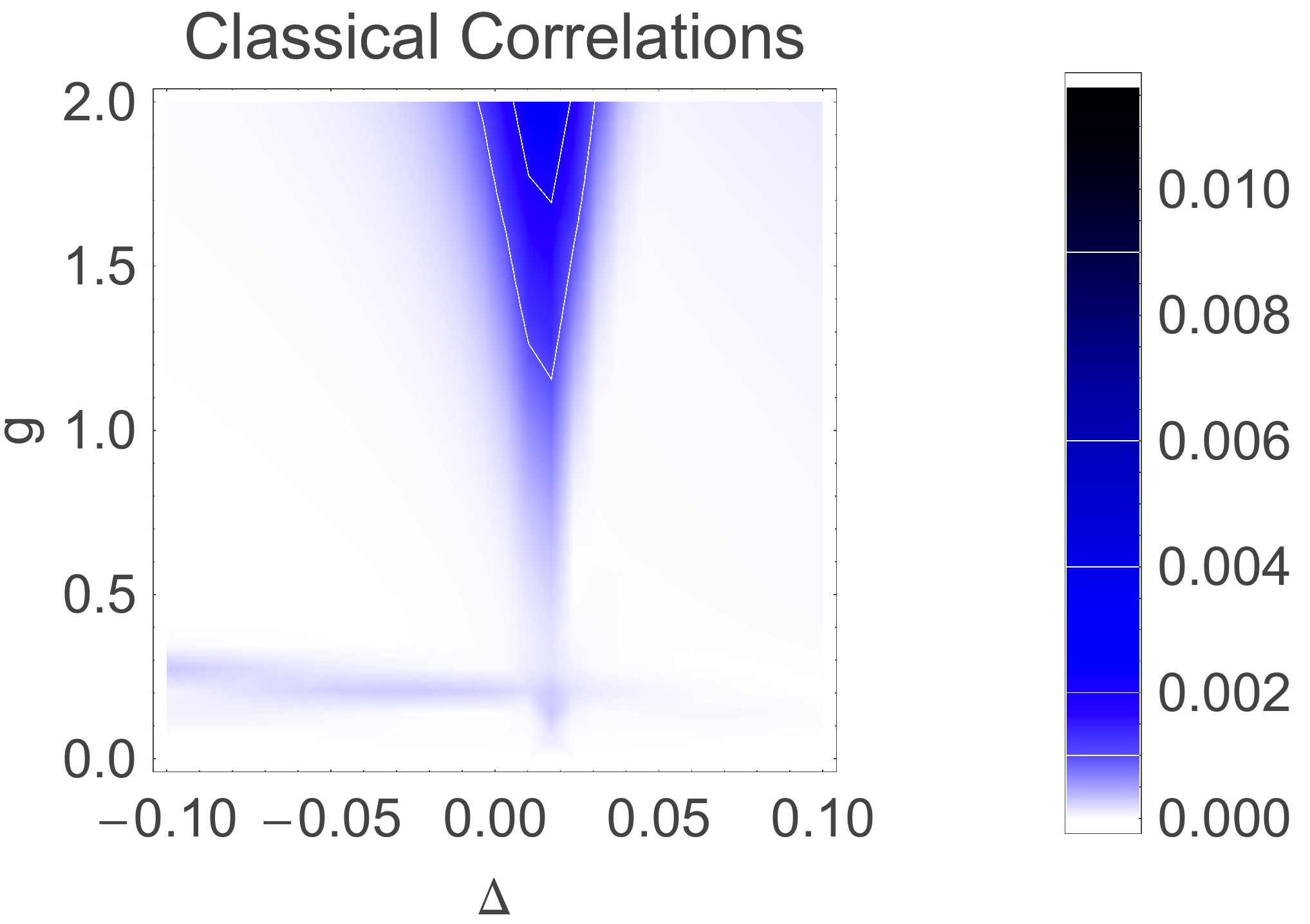}  \includegraphics[width= 0.75 \columnwidth] {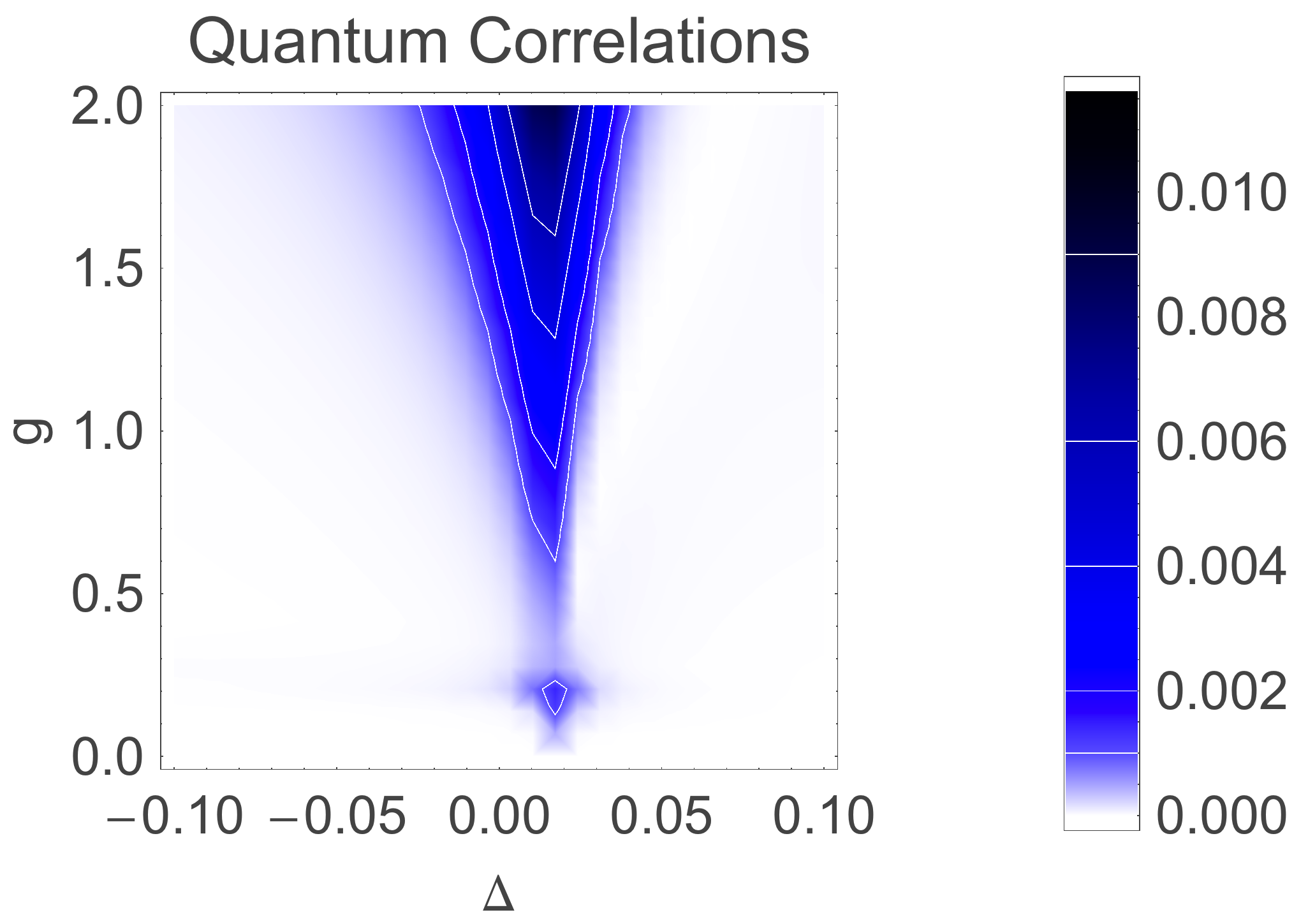}
 \caption{Classical information (top) and quantum discord (bottom) between the two qubits as a function of the optical coupling constant $g$ and detuning $\Delta$. The other parameters are as in Fig.\ \ref{mu100}. }
 \label{fig:class-discord}
\end{figure}

The results are shown in figure~\ref{fig:class-discord}. We find that the classical information and the quantum discord give almost identical plots (apart from different numerical scales), recovering the same behavior observed by the quantum mutual information. Therefore, we can say the synchronization process is responsible for the creation of both classical and quantum correlations in the system. Actually, in this particular example, the amount of quantum correlations appears to be the dominant contribution to the mutual information and this fact could be associated to the deep quantum nature of the system (two qubits) as already observed in other contexts \cite{Giorgi2012a,Giorgi2013a}. In general we expect the interplay between classical and quantum correlations to strongly depend on the particular system under investigation. 

We also comment that, in all the parameter space that we explored in the analysis of the two-qubit model, we did not found any entanglement. This is consistent with the well established knowledge that entanglement is a much stronger form of correlation with respect to quantum discord. So one can expect that it is difficult to generate entanglement by exploiting the effect of spontaneous synchronization. This difficulty has been already noticed in \cite{Mari2013a} but, nonetheless, other works found significant amounts of entanglement \cite{Manzano2013a, Lee2014a} in different systems subject to synchronization.  Again, as for the case of quantum correlations,  we can conclude that also the relationship between entanglement and synchronization strongly depends on the specific details of system.  

\section{Conclusions}

In this work we introduced a new approach to the analysis of synchronization effects in quantum mechanical systems. Specifically we proposed the use of quantum mutual information as an order parameter for signalling the presence or the absence of quantum synchronization. With respect to other specific or semiclassical measures of synchronization, mutual information is well defined for every bipartite quantum state and does not depend on the particular details of the system. Indeed in this work we have been able to analyse, within the same theoretical framework,  the  synchronization of completely different devices: namely two quantum Van der Pol resonators and two qubits. Given the universality of the concept of information, we expect that our approach could be successfully applied also to other systems like: non-linear optical cavities \cite{Lee2013b,Jiasen},  opto-mechanical arrays \cite{Ludwig2013a}.

\section{Acknowledgements}
This work was supported by  ERC-Advanced SouLMan (Grant Agreement 321122), EU collaborative Project TherMiQ (Grant Agreement 618074), EU-IP-SIQS (Grant Agreement 600645), by Italian MIUR via PRIN Project 2010LLKJBX and by SNS-Project "Non-equilibrium dynamics of one-dimensional quantum systems".

\bibliography{sci}

\begin{thebibliography}{28}%
\makeatletter
\providecommand \@ifxundefined [1]{%
 \@ifx{#1\undefined}
}%
\providecommand \@ifnum [1]{%
 \ifnum #1\expandafter \@firstoftwo
 \else \expandafter \@secondoftwo
 \fi
}%
\providecommand \@ifx [1]{%
 \ifx #1\expandafter \@firstoftwo
 \else \expandafter \@secondoftwo
 \fi
}%
\providecommand \natexlab [1]{#1}%
\providecommand \enquote  [1]{``#1''}%
\providecommand \bibnamefont  [1]{#1}%
\providecommand \bibfnamefont [1]{#1}%
\providecommand \citenamefont [1]{#1}%
\providecommand \href@noop [0]{\@secondoftwo}%
\providecommand \href [0]{\begingroup \@sanitize@url \@href}%
\providecommand \@href[1]{\@@startlink{#1}\@@href}%
\providecommand \@@href[1]{\endgroup#1\@@endlink}%
\providecommand \@sanitize@url [0]{\catcode `\\12\catcode `\$12\catcode
  `\&12\catcode `\#12\catcode `\^12\catcode `\_12\catcode `\%12\relax}%
\providecommand \@@startlink[1]{}%
\providecommand \@@endlink[0]{}%
\providecommand \url  [0]{\begingroup\@sanitize@url \@url }%
\providecommand \@url [1]{\endgroup\@href {#1}{\urlprefix }}%
\providecommand \urlprefix  [0]{URL }%
\providecommand \Eprint [0]{\href }%
\providecommand \doibase [0]{http://dx.doi.org/}%
\providecommand \selectlanguage [0]{\@gobble}%
\providecommand \bibinfo  [0]{\@secondoftwo}%
\providecommand \bibfield  [0]{\@secondoftwo}%
\providecommand \translation [1]{[#1]}%
\providecommand \BibitemOpen [0]{}%
\providecommand \bibitemStop [0]{}%
\providecommand \bibitemNoStop [0]{.\EOS\space}%
\providecommand \EOS [0]{\spacefactor3000\relax}%
\providecommand \BibitemShut  [1]{\csname bibitem#1\endcsname}%
\let\auto@bib@innerbib\@empty
\bibitem [{\citenamefont {Huygens}()}]{huygens5letterbis}%
  \BibitemOpen
  \bibfield  {author} {\bibinfo {author} {\bibfnamefont {C.}~\bibnamefont
  {Huygens}},\ }\href@noop {} {\emph {\bibinfo {title} {Oeuvres Compl{\`e}te de
  Christiaan Huygens: Correspondence}}},\ Vol.~\bibinfo {volume} {5},\ p.\
  \bibinfo {pages} {246}\BibitemShut {NoStop}%
\bibitem [{\citenamefont {Pikovsky}\ \emph {et~al.}(2003)\citenamefont
  {Pikovsky}, \citenamefont {Rosenblum},\ and\ \citenamefont
  {Kurths}}]{pikovsky}%
  \BibitemOpen
  \bibfield  {author} {\bibinfo {author} {\bibfnamefont {A.}~\bibnamefont
  {Pikovsky}}, \bibinfo {author} {\bibfnamefont {M.}~\bibnamefont {Rosenblum}},
  \ and\ \bibinfo {author} {\bibfnamefont {J.}~\bibnamefont {Kurths}},\
  }\href@noop {} {\emph {\bibinfo {title} {Synchronization: A universal concept
  in nonlinear sciences}}},\ Vol.~\bibinfo {volume} {12}\ (\bibinfo
  {publisher} {Cambridge University Press},\ \bibinfo {year}
  {2003})\BibitemShut {NoStop}%
\bibitem [{\citenamefont {Strogatz}(2001)}]{strogatz}%
  \BibitemOpen
  \bibfield  {author} {\bibinfo {author} {\bibfnamefont {S.~H.}\ \bibnamefont
  {Strogatz}},\ }\href@noop {} {\emph {\bibinfo {title} {Nonlinear dynamics and
  chaos: with applications to physics, biology and chemistry}}}\ (\bibinfo
  {publisher} {Perseus publishing},\ \bibinfo {year} {2001})\BibitemShut
  {NoStop}%
\bibitem [{\citenamefont {Lee}\ and\ \citenamefont {Cross}(2013)}]{Lee2013b}%
  \BibitemOpen
  \bibfield  {author} {\bibinfo {author} {\bibfnamefont {T.~E.}\ \bibnamefont
  {Lee}}\ and\ \bibinfo {author} {\bibfnamefont {M.~C.}\ \bibnamefont
  {Cross}},\ }\href {\doibase 10.1103/PhysRevA.88.013834} {\bibfield  {journal}
  {\bibinfo  {journal} {Phys. Rev. A}\ }\textbf {\bibinfo {volume} {88}},\
  \bibinfo {pages} {013834} (\bibinfo {year} {2013})}\BibitemShut {NoStop}%
\bibitem [{\citenamefont {Heinrich}\ \emph {et~al.}(2011)\citenamefont
  {Heinrich}, \citenamefont {Ludwig}, \citenamefont {Qian}, \citenamefont
  {Kubala},\ and\ \citenamefont {Marquardt}}]{Heinrich2011a}%
  \BibitemOpen
  \bibfield  {author} {\bibinfo {author} {\bibfnamefont {G.}~\bibnamefont
  {Heinrich}}, \bibinfo {author} {\bibfnamefont {M.}~\bibnamefont {Ludwig}},
  \bibinfo {author} {\bibfnamefont {J.}~\bibnamefont {Qian}}, \bibinfo {author}
  {\bibfnamefont {B.}~\bibnamefont {Kubala}}, \ and\ \bibinfo {author}
  {\bibfnamefont {F.}~\bibnamefont {Marquardt}},\ }\href {\doibase
  10.1103/PhysRevLett.107.043603} {\bibfield  {journal} {\bibinfo  {journal}
  {Phys. Rev. Lett.}\ }\textbf {\bibinfo {volume} {107}},\ \bibinfo {pages}
  {043603} (\bibinfo {year} {2011})}\BibitemShut {NoStop}%
\bibitem [{\citenamefont {Ludwig}\ and\ \citenamefont
  {Marquardt}(2013)}]{Ludwig2013a}%
  \BibitemOpen
  \bibfield  {author} {\bibinfo {author} {\bibfnamefont {M.}~\bibnamefont
  {Ludwig}}\ and\ \bibinfo {author} {\bibfnamefont {F.}~\bibnamefont
  {Marquardt}},\ }\href {\doibase 10.1103/PhysRevLett.111.073603} {\bibfield
  {journal} {\bibinfo  {journal} {Phys. Rev. Lett.}\ }\textbf {\bibinfo
  {volume} {111}},\ \bibinfo {pages} {073603} (\bibinfo {year}
  {2013})}\BibitemShut {NoStop}%
\bibitem [{\citenamefont {Mari}\ \emph {et~al.}(2013)\citenamefont {Mari},
  \citenamefont {Farace}, \citenamefont {Didier}, \citenamefont {Giovannetti},\
  and\ \citenamefont {Fazio}}]{Mari2013a}%
  \BibitemOpen
  \bibfield  {author} {\bibinfo {author} {\bibfnamefont {A.}~\bibnamefont
  {Mari}}, \bibinfo {author} {\bibfnamefont {A.}~\bibnamefont {Farace}},
  \bibinfo {author} {\bibfnamefont {N.}~\bibnamefont {Didier}}, \bibinfo
  {author} {\bibfnamefont {V.}~\bibnamefont {Giovannetti}}, \ and\ \bibinfo
  {author} {\bibfnamefont {R.}~\bibnamefont {Fazio}},\ }\href {\doibase
  10.1103/PhysRevLett.111.103605} {\bibfield  {journal} {\bibinfo  {journal}
  {Phys. Rev. Lett.}\ }\textbf {\bibinfo {volume} {111}},\ \bibinfo {pages}
  {103605} (\bibinfo {year} {2013})}\BibitemShut {NoStop}%
\bibitem [{\citenamefont {Lee}\ and\ \citenamefont
  {Sadeghpour}(2013)}]{Lee2013a}%
  \BibitemOpen
  \bibfield  {author} {\bibinfo {author} {\bibfnamefont {T.~E.}\ \bibnamefont
  {Lee}}\ and\ \bibinfo {author} {\bibfnamefont {H.~R.}\ \bibnamefont
  {Sadeghpour}},\ }\href {\doibase 10.1103/PhysRevLett.111.234101} {\bibfield
  {journal} {\bibinfo  {journal} {Phys. Rev. Lett.}\ }\textbf {\bibinfo
  {volume} {111}},\ \bibinfo {pages} {234101} (\bibinfo {year}
  {2013})}\BibitemShut {NoStop}%
\bibitem [{\citenamefont {Lee}\ \emph {et~al.}(2014)\citenamefont {Lee},
  \citenamefont {Chan},\ and\ \citenamefont {Wang}}]{Lee2014a}%
  \BibitemOpen
  \bibfield  {author} {\bibinfo {author} {\bibfnamefont {T.~E.}\ \bibnamefont
  {Lee}}, \bibinfo {author} {\bibfnamefont {C.-K.}\ \bibnamefont {Chan}}, \
  and\ \bibinfo {author} {\bibfnamefont {S.}~\bibnamefont {Wang}},\ }\href
  {\doibase 10.1103/PhysRevE.89.022913} {\bibfield  {journal} {\bibinfo
  {journal} {Phys. Rev. E}\ }\textbf {\bibinfo {volume} {89}},\ \bibinfo
  {pages} {022913} (\bibinfo {year} {2014})}\BibitemShut {NoStop}%
\bibitem [{\citenamefont {Walter}\ \emph
  {et~al.}(2014{\natexlab{a}})\citenamefont {Walter}, \citenamefont
  {Nunnenkamp},\ and\ \citenamefont {Bruder}}]{Walter2014b}%
  \BibitemOpen
  \bibfield  {author} {\bibinfo {author} {\bibfnamefont {S.}~\bibnamefont
  {Walter}}, \bibinfo {author} {\bibfnamefont {A.}~\bibnamefont {Nunnenkamp}},
  \ and\ \bibinfo {author} {\bibfnamefont {C.}~\bibnamefont {Bruder}},\ }\href
  {\doibase 10.1002/andp.201400144} {\bibfield  {journal} {\bibinfo  {journal}
  {Annalen der Physik}\ } (\bibinfo {year} {2014}{\natexlab{a}}),\
  10.1002/andp.201400144}\BibitemShut {NoStop}%
\bibitem [{\citenamefont {Walter}\ \emph
  {et~al.}(2014{\natexlab{b}})\citenamefont {Walter}, \citenamefont
  {Nunnenkamp},\ and\ \citenamefont {Bruder}}]{Walter2014a}%
  \BibitemOpen
  \bibfield  {author} {\bibinfo {author} {\bibfnamefont {S.}~\bibnamefont
  {Walter}}, \bibinfo {author} {\bibfnamefont {A.}~\bibnamefont {Nunnenkamp}},
  \ and\ \bibinfo {author} {\bibfnamefont {C.}~\bibnamefont {Bruder}},\ }\href
  {\doibase 10.1103/PhysRevLett.112.094102} {\bibfield  {journal} {\bibinfo
  {journal} {Phys. Rev. Lett.}\ }\textbf {\bibinfo {volume} {112}},\ \bibinfo
  {pages} {094102} (\bibinfo {year} {2014}{\natexlab{b}})}\BibitemShut
  {NoStop}%
\bibitem [{\citenamefont {Shim}\ \emph {et~al.}(2007)\citenamefont {Shim},
  \citenamefont {Imboden},\ and\ \citenamefont {Mohanty}}]{Shim2007}%
  \BibitemOpen
  \bibfield  {author} {\bibinfo {author} {\bibfnamefont {S.-B.}\ \bibnamefont
  {Shim}}, \bibinfo {author} {\bibfnamefont {M.}~\bibnamefont {Imboden}}, \
  and\ \bibinfo {author} {\bibfnamefont {P.}~\bibnamefont {Mohanty}},\ }\href
  {\doibase 10.1126/science.1137307} {\bibfield  {journal} {\bibinfo  {journal}
  {Science}\ }\textbf {\bibinfo {volume} {316}},\ \bibinfo {pages} {95}
  (\bibinfo {year} {2007})}\BibitemShut {NoStop}%
\bibitem [{\citenamefont {Zhang}\ \emph {et~al.}(2012)\citenamefont {Zhang},
  \citenamefont {Wiederhecker}, \citenamefont {Manipatruni}, \citenamefont
  {Barnard}, \citenamefont {McEuen},\ and\ \citenamefont {Lipson}}]{Zhang2012}%
  \BibitemOpen
  \bibfield  {author} {\bibinfo {author} {\bibfnamefont {M.}~\bibnamefont
  {Zhang}}, \bibinfo {author} {\bibfnamefont {G.~S.}\ \bibnamefont
  {Wiederhecker}}, \bibinfo {author} {\bibfnamefont {S.}~\bibnamefont
  {Manipatruni}}, \bibinfo {author} {\bibfnamefont {A.}~\bibnamefont
  {Barnard}}, \bibinfo {author} {\bibfnamefont {P.}~\bibnamefont {McEuen}}, \
  and\ \bibinfo {author} {\bibfnamefont {M.}~\bibnamefont {Lipson}},\ }\href
  {\doibase 10.1103/PhysRevLett.109.233906} {\bibfield  {journal} {\bibinfo
  {journal} {Phys. Rev. Lett.}\ }\textbf {\bibinfo {volume} {109}},\ \bibinfo
  {pages} {233906} (\bibinfo {year} {2012})}\BibitemShut {NoStop}%
\bibitem [{\citenamefont {Bagheri}\ \emph {et~al.}(2013)\citenamefont
  {Bagheri}, \citenamefont {Poot}, \citenamefont {Fan}, \citenamefont
  {Marquardt},\ and\ \citenamefont {Tang}}]{Bagheri2013a}%
  \BibitemOpen
  \bibfield  {author} {\bibinfo {author} {\bibfnamefont {M.}~\bibnamefont
  {Bagheri}}, \bibinfo {author} {\bibfnamefont {M.}~\bibnamefont {Poot}},
  \bibinfo {author} {\bibfnamefont {L.}~\bibnamefont {Fan}}, \bibinfo {author}
  {\bibfnamefont {F.}~\bibnamefont {Marquardt}}, \ and\ \bibinfo {author}
  {\bibfnamefont {H.~X.}\ \bibnamefont {Tang}},\ }\href {\doibase
  10.1103/PhysRevLett.111.213902} {\bibfield  {journal} {\bibinfo  {journal}
  {Phys. Rev. Lett.}\ }\textbf {\bibinfo {volume} {111}},\ \bibinfo {pages}
  {213902} (\bibinfo {year} {2013})}\BibitemShut {NoStop}%
\bibitem [{\citenamefont {Matheny}\ \emph {et~al.}(2014)\citenamefont
  {Matheny}, \citenamefont {Grau}, \citenamefont {Villanueva}, \citenamefont
  {Karabalin}, \citenamefont {Cross},\ and\ \citenamefont
  {Roukes}}]{Matheny2014}%
  \BibitemOpen
  \bibfield  {author} {\bibinfo {author} {\bibfnamefont {M.~H.}\ \bibnamefont
  {Matheny}}, \bibinfo {author} {\bibfnamefont {M.}~\bibnamefont {Grau}},
  \bibinfo {author} {\bibfnamefont {L.~G.}\ \bibnamefont {Villanueva}},
  \bibinfo {author} {\bibfnamefont {R.~B.}\ \bibnamefont {Karabalin}}, \bibinfo
  {author} {\bibfnamefont {M.~C.}\ \bibnamefont {Cross}}, \ and\ \bibinfo
  {author} {\bibfnamefont {M.~L.}\ \bibnamefont {Roukes}},\ }\href {\doibase
  10.1103/PhysRevLett.112.014101} {\bibfield  {journal} {\bibinfo  {journal}
  {Phys. Rev. Lett.}\ }\textbf {\bibinfo {volume} {112}},\ \bibinfo {pages}
  {014101} (\bibinfo {year} {2014})}\BibitemShut {NoStop}%
\bibitem [{\citenamefont {Giorgi}\ \emph {et~al.}(2012)\citenamefont {Giorgi},
  \citenamefont {Galve}, \citenamefont {Manzano}, \citenamefont {Colet},\ and\
  \citenamefont {Zambrini}}]{Giorgi2012a}%
  \BibitemOpen
  \bibfield  {author} {\bibinfo {author} {\bibfnamefont {G.~L.}\ \bibnamefont
  {Giorgi}}, \bibinfo {author} {\bibfnamefont {F.}~\bibnamefont {Galve}},
  \bibinfo {author} {\bibfnamefont {G.}~\bibnamefont {Manzano}}, \bibinfo
  {author} {\bibfnamefont {P.}~\bibnamefont {Colet}}, \ and\ \bibinfo {author}
  {\bibfnamefont {R.}~\bibnamefont {Zambrini}},\ }\href {\doibase
  10.1103/PhysRevA.85.052101} {\bibfield  {journal} {\bibinfo  {journal} {Phys.
  Rev. A}\ }\textbf {\bibinfo {volume} {85}},\ \bibinfo {pages} {052101}
  (\bibinfo {year} {2012})}\BibitemShut {NoStop}%
\bibitem [{\citenamefont {Manzano}\ \emph {et~al.}(2013)\citenamefont
  {Manzano}, \citenamefont {Galve}, \citenamefont {Giorgi}, \citenamefont
  {Hernandez-Garcia},\ and\ \citenamefont {Zambrini}}]{Manzano2013a}%
  \BibitemOpen
  \bibfield  {author} {\bibinfo {author} {\bibfnamefont {G.}~\bibnamefont
  {Manzano}}, \bibinfo {author} {\bibfnamefont {F.}~\bibnamefont {Galve}},
  \bibinfo {author} {\bibfnamefont {G.~L.}\ \bibnamefont {Giorgi}}, \bibinfo
  {author} {\bibfnamefont {E.}~\bibnamefont {Hernandez-Garcia}}, \ and\
  \bibinfo {author} {\bibfnamefont {R.}~\bibnamefont {Zambrini}},\ }\href
  {http://dx.doi.org/10.1038/srep01439} {\bibfield  {journal} {\bibinfo
  {journal} {Sci. Rep.}\ }\textbf {\bibinfo {volume} {3}} (\bibinfo {year}
  {2013})}\BibitemShut {NoStop}%
\bibitem [{\citenamefont {Modi}\ \emph {et~al.}(2012)\citenamefont {Modi},
  \citenamefont {Brodutch}, \citenamefont {Cable}, \citenamefont {Paterek},\
  and\ \citenamefont {Vedral}}]{Modi2012a}%
  \BibitemOpen
  \bibfield  {author} {\bibinfo {author} {\bibfnamefont {K.}~\bibnamefont
  {Modi}}, \bibinfo {author} {\bibfnamefont {A.}~\bibnamefont {Brodutch}},
  \bibinfo {author} {\bibfnamefont {H.}~\bibnamefont {Cable}}, \bibinfo
  {author} {\bibfnamefont {T.}~\bibnamefont {Paterek}}, \ and\ \bibinfo
  {author} {\bibfnamefont {V.}~\bibnamefont {Vedral}},\ }\href {\doibase
  10.1103/RevModPhys.84.1655} {\bibfield  {journal} {\bibinfo  {journal} {Rev.
  Mod. Phys.}\ }\textbf {\bibinfo {volume} {84}},\ \bibinfo {pages} {1655}
  (\bibinfo {year} {2012})}\BibitemShut {NoStop}%
\bibitem [{\citenamefont {Giorgi}\ \emph {et~al.}(2013)\citenamefont {Giorgi},
  \citenamefont {Plastina}, \citenamefont {Francica},\ and\ \citenamefont
  {Zambrini}}]{Giorgi2013a}%
  \BibitemOpen
  \bibfield  {author} {\bibinfo {author} {\bibfnamefont {G.~L.}\ \bibnamefont
  {Giorgi}}, \bibinfo {author} {\bibfnamefont {F.}~\bibnamefont {Plastina}},
  \bibinfo {author} {\bibfnamefont {G.}~\bibnamefont {Francica}}, \ and\
  \bibinfo {author} {\bibfnamefont {R.}~\bibnamefont {Zambrini}},\ }\href
  {\doibase 10.1103/PhysRevA.88.042115} {\bibfield  {journal} {\bibinfo
  {journal} {Phys. Rev. A}\ }\textbf {\bibinfo {volume} {88}},\ \bibinfo
  {pages} {042115} (\bibinfo {year} {2013})}\BibitemShut {NoStop}%
\bibitem [{\citenamefont {Qiu}\ \emph {et~al.}(2014)\citenamefont {Qiu},
  \citenamefont {Juli\'a-D\'iaz}, \citenamefont {Garcia-March},\ and\
  \citenamefont {Polls}}]{Qiu2014a}%
  \BibitemOpen
  \bibfield  {author} {\bibinfo {author} {\bibfnamefont {H.}~\bibnamefont
  {Qiu}}, \bibinfo {author} {\bibfnamefont {B.}~\bibnamefont {Juli\'a-D\'iaz}},
  \bibinfo {author} {\bibfnamefont {M.~A.}\ \bibnamefont {Garcia-March}}, \
  and\ \bibinfo {author} {\bibfnamefont {A.}~\bibnamefont {Polls}},\ }\href
  {\doibase 10.1103/PhysRevA.90.033603} {\bibfield  {journal} {\bibinfo
  {journal} {Phys. Rev. A}\ }\textbf {\bibinfo {volume} {90}},\ \bibinfo
  {pages} {033603} (\bibinfo {year} {2014})}\BibitemShut {NoStop}%
\bibitem [{\citenamefont {Xu}\ \emph {et~al.}(2014)\citenamefont {Xu},
  \citenamefont {Tieri}, \citenamefont {Fine}, \citenamefont {Thompson},\ and\
  \citenamefont {Holland}}]{tieri}%
  \BibitemOpen
  \bibfield  {author} {\bibinfo {author} {\bibfnamefont {M.}~\bibnamefont
  {Xu}}, \bibinfo {author} {\bibfnamefont {D.~A.}\ \bibnamefont {Tieri}},
  \bibinfo {author} {\bibfnamefont {E.~C.}\ \bibnamefont {Fine}}, \bibinfo
  {author} {\bibfnamefont {J.~K.}\ \bibnamefont {Thompson}}, \ and\ \bibinfo
  {author} {\bibfnamefont {M.~J.}\ \bibnamefont {Holland}},\ }\href {\doibase
  10.1103/PhysRevLett.113.154101} {\bibfield  {journal} {\bibinfo  {journal}
  {Phys. Rev. Lett.}\ }\textbf {\bibinfo {volume} {113}},\ \bibinfo {pages}
  {154101} (\bibinfo {year} {2014})}\BibitemShut {NoStop}%
\bibitem [{\citenamefont {Van~der Pol}(1920)}]{van1934nonlinear}%
  \BibitemOpen
  \bibfield  {author} {\bibinfo {author} {\bibfnamefont {B.}~\bibnamefont
  {Van~der Pol}},\ }\href@noop {} {\bibfield  {journal} {\bibinfo  {journal}
  {Radio Review}\ }\textbf {\bibinfo {volume} {1}} (\bibinfo {year}
  {1920})}\BibitemShut {NoStop}%
\bibitem [{\citenamefont {Henderson}\ and\ \citenamefont
  {Vedral}(2001)}]{Henderson2001a}%
  \BibitemOpen
  \bibfield  {author} {\bibinfo {author} {\bibfnamefont {L.}~\bibnamefont
  {Henderson}}\ and\ \bibinfo {author} {\bibfnamefont {V.}~\bibnamefont
  {Vedral}},\ }\href {\doibase 10.1088/0305-4470/34/35/315} {\bibfield
  {journal} {\bibinfo  {journal} {Journal of Physics A: Mathematical and
  General}\ }\textbf {\bibinfo {volume} {34}},\ \bibinfo {pages} {6899}
  (\bibinfo {year} {2001})}\BibitemShut {NoStop}%
\bibitem [{\citenamefont {Ollivier}\ and\ \citenamefont
  {Zurek}(2001)}]{Ollivier2001a}%
  \BibitemOpen
  \bibfield  {author} {\bibinfo {author} {\bibfnamefont {H.}~\bibnamefont
  {Ollivier}}\ and\ \bibinfo {author} {\bibfnamefont {W.~H.}\ \bibnamefont
  {Zurek}},\ }\href {\doibase 10.1103/PhysRevLett.88.017901} {\bibfield
  {journal} {\bibinfo  {journal} {Phys. Rev. Lett.}\ }\textbf {\bibinfo
  {volume} {88}},\ \bibinfo {pages} {017901} (\bibinfo {year}
  {2001})}\BibitemShut {NoStop}%
\bibitem [{\citenamefont {Hamieh}\ \emph {et~al.}(2004)\citenamefont {Hamieh},
  \citenamefont {Kobes},\ and\ \citenamefont {Zaraket}}]{Hamieh2004a}%
  \BibitemOpen
  \bibfield  {author} {\bibinfo {author} {\bibfnamefont {S.}~\bibnamefont
  {Hamieh}}, \bibinfo {author} {\bibfnamefont {R.}~\bibnamefont {Kobes}}, \
  and\ \bibinfo {author} {\bibfnamefont {H.}~\bibnamefont {Zaraket}},\ }\href
  {\doibase 10.1103/PhysRevA.70.052325} {\bibfield  {journal} {\bibinfo
  {journal} {Phys. Rev. A}\ }\textbf {\bibinfo {volume} {70}},\ \bibinfo
  {pages} {052325} (\bibinfo {year} {2004})}\BibitemShut {NoStop}%
\bibitem [{\citenamefont {D'Ariano}\ \emph {et~al.}(2005)\citenamefont
  {D'Ariano}, \citenamefont {Lo~Presti},\ and\ \citenamefont
  {Perinotti}}]{DAriano2005a}%
  \BibitemOpen
  \bibfield  {author} {\bibinfo {author} {\bibfnamefont {G.~M.}\ \bibnamefont
  {D'Ariano}}, \bibinfo {author} {\bibfnamefont {P.}~\bibnamefont {Lo~Presti}},
  \ and\ \bibinfo {author} {\bibfnamefont {P.}~\bibnamefont {Perinotti}},\
  }\href {\doibase 10.1088/0305-4470/38/26/010} {\bibfield  {journal} {\bibinfo
   {journal} {Journal of Physics A: Mathematical and General}\ }\textbf
  {\bibinfo {volume} {38}},\ \bibinfo {pages} {5979} (\bibinfo {year}
  {2005})}\BibitemShut {NoStop}%
\bibitem [{\citenamefont {Galve}\ \emph {et~al.}(2011)\citenamefont {Galve},
  \citenamefont {Giorgi},\ and\ \citenamefont {Zambrini}}]{Galve2011a}%
  \BibitemOpen
  \bibfield  {author} {\bibinfo {author} {\bibfnamefont {F.}~\bibnamefont
  {Galve}}, \bibinfo {author} {\bibfnamefont {G.~L.}\ \bibnamefont {Giorgi}}, \
  and\ \bibinfo {author} {\bibfnamefont {R.}~\bibnamefont {Zambrini}},\ }\href
  {\doibase doi:10.1209/0295-5075/96/40005} {\bibfield  {journal} {\bibinfo
  {journal} {EPL (Europhysics Letters)}\ }\textbf {\bibinfo {volume} {96}},\
  \bibinfo {pages} {40005} (\bibinfo {year} {2011})}\BibitemShut {NoStop}%
\bibitem [{\citenamefont {Jin}\ \emph {et~al.}(2013)\citenamefont {Jin},
  \citenamefont {Rossini}, \citenamefont {Fazio}, \citenamefont {Leib},\ and\
  \citenamefont {Hartmann}}]{Jiasen}%
  \BibitemOpen
  \bibfield  {author} {\bibinfo {author} {\bibfnamefont {J.}~\bibnamefont
  {Jin}}, \bibinfo {author} {\bibfnamefont {D.}~\bibnamefont {Rossini}},
  \bibinfo {author} {\bibfnamefont {R.}~\bibnamefont {Fazio}}, \bibinfo
  {author} {\bibfnamefont {M.}~\bibnamefont {Leib}}, \ and\ \bibinfo {author}
  {\bibfnamefont {M.~J.}\ \bibnamefont {Hartmann}},\ }\href {\doibase
  10.1103/PhysRevLett.110.163605} {\bibfield  {journal} {\bibinfo  {journal}
  {Phys. Rev. Lett.}\ }\textbf {\bibinfo {volume} {110}},\ \bibinfo {pages}
  {163605} (\bibinfo {year} {2013})}\BibitemShut {NoStop}%
\end{thebibliography}%
\bibliographystyle{apsrev4-1} 

\end{document}